\documentclass[reprint, notitlepage, twocolumn,amsmath,amssymb, aps,prl]{revtex4-1}
\usepackage{amsmath}
\usepackage{graphicx}
\usepackage{color}
\usepackage[colorlinks,citecolor=magenta]{hyperref}

\newcommand{\kt}[1]{\left|#1\right>}

\newcommand{\brkt}[2]{\left<#1|#2 \right>}

\newcommand{\me}[3]{\left<#1\right|#2\left|#3 \right>}
\newcommand{\mesq}[3]{\left| \left<#1\right|#2\left|#3 \right>\right|^2}

\begin{document}
\title{Strong exciton regulation of Raman scattering in monolayer dichalcogenides}
\author{Yuanxi Wang}
\email{yow5110@psu.edu}
\affiliation{Department of Physics, Pennsylvania State University, University Park, Pennsylvania 16802, USA}
\affiliation{2-Dimensional Crystal Consortium,Pennsylvania State University, University Park, Pennsylvania 16802}
\affiliation{Material Research Institute, Pennsylvania State University, University Park, Pennsylvania 16802}

\author{Bruno R. Carvalho} 
\affiliation{Departamento de F\'isica, Universidade Federal do Rio Grande do Norte, Natal, Rio Grande do Norte 59078-970, Brazil}

\author{Vincent H. Crespi} 
\email{vhc2@psu.edu}
\affiliation{Department of Physics, Pennsylvania State University, University Park, Pennsylvania 16802, USA}
\affiliation{\mbox{Department of Materials Science and Engineering, Pennsylvania State University, University Park, Pennsylvania 16802, USA}}
\affiliation{Department of Chemistry, Pennsylvania State University, University Park, Pennsylvania 16802, USA}
\affiliation{2-Dimensional Crystal Consortium,Pennsylvania State University, University Park, Pennsylvania 16802}

\begin{abstract}
The weakly screened electron-hole interactions in an atomically thin semiconductor not only downshift its excitation spectrum from a quasiparticle one, but also redistribute excitation energies and wavefunction characters with profound effects on diverse modes of material response, including the exciton-phonon scattering processes accessible to resonant Raman measurements. Here we develop a first-principles framework to calculate frequency-dependent resonant Raman intensities that includes excitonic effects and goes beyond the Placzek approximation. We show how excitonic effects in MoS$_2$ strongly regulate Raman scattering amplitudes and thereby explain the puzzling near-absence of resonant Raman response around the A and B excitons (which produce very strong signals in optical absorption), and also the pronounced strength of the resonant Raman response from the C exciton. Furthermore, this efficient perturbative approach reduces the number of GW-BSE calculations from two per Raman mode (in finite displacement) to one for all modes and affords natural extension to higher-order resonant Raman processes.
\end{abstract}

\maketitle
Low-energy excitations of two-dimensional semiconductors such as MoS$_2$ are dominated by very strong excitonic effects \cite{Qiu2013, Soklaski2014}. While excitonic resonances are evident from absorption spectroscopy \cite{Mak2010}, resonant Raman spectroscopy offers a more multifaceted perspective: the Raman intensity of a phonon mode plotted against laser energies (Raman excitation profile) not only reveals excitonic resonances with resolutions on par with absorption, but also reflects exciton-phonon coupling strengths \cite{Carvalho2015, Carvalho2017}.  Raman features emerging upon reaching resonance furthermore capture finite-momentum processes such as higher-order Raman scattering and defect scattering \cite{Lee2015, Golasa2014, Mignuzzi2015, Ferrari2006, Venezuela2011b}, both key processes in valleytronics \cite{Carvalho2017, Zeng2012, Mak2012}. The appeal of these rich outputs, combined with the procedural simplicity of Raman measurements (perhaps best attested by Raman's original discovery using sunlight, distilled liquids, and a human eye \cite{Raman1928}) contribute to its wide usage. Spectral features in Raman excitation profiles are generally aligned in energy with  absorption features for molecules \cite{Asher1986, Shelnutt1980, Page1991} and 3D bulk solids \cite{Martin1983}, with intensities of similar orders of magnitude, as modulated by electron-phonon interactions. This expectation is confounded by the puzzling near-absence of Raman intensity measured at the A/B exciton energies and the disproportionately higher Raman intensity measured at the C exciton in MoS$_2$ \cite{Carvalho2016, Lee2015}. This anomaly in 2D semiconductors suggests an unidentified regulating effect by excitons.

Despite the rich experimental data on Raman measurements of 2D solids, the role of excitons on Raman spectra is rarely modeled at a first-principles level beyond calculating shifted resonance energies, because of the high computational cost of many-body perturbation theory calculations and the sparsity of implementations that consolidate electron-phonon and many-body phenomena. 
One recent important theoretical advance \cite{Gillet2013} (implemented in \cite{DelCorro2016, Miranda2017}) used finite differences through solution of the Bethe-Salpeter equation (BSE) on the quasiparticle (GW) band structure, but employed a quasi-static Placzek approximation that is only valid in the non-resonant regime. Here we follow the generalized approach of Ref.~\cite{Profeta2001}, valid for solids in the resonant regime, to develop a perturbation framework that goes beyond the Placzek approximation and includes electron-hole interactions. Both ingredients are crucial to capturing exciton-regulated Raman scattering in MoS$_2$, including dramatic differences in the strength of the resonant response in the vicinity of the A/B and C excitons that agree with experiments.
We show that band-extrema electron-hole pairs such as the A/B excitons suppress Raman response due to their energies being well separated from the rest of the exciton spectrum, and that parallel-band electron-hole pairs such as the C exciton amplify Raman response due to their bunching of energies causing strong rehybridization during atomic vibration.   

\begin{figure}[h]
\centering
\includegraphics[width=0.5\textwidth]{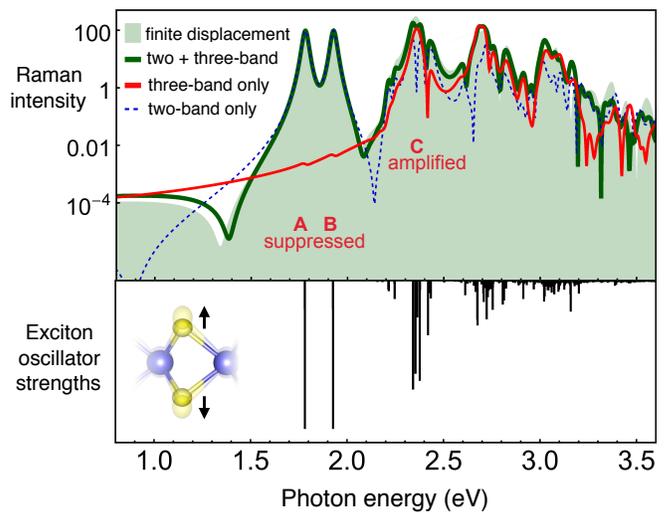}
\caption{Resonant Raman intensities of the out-of-plane $A'_1$ mode in MoS$_2$ calculated as a function of the laser energy. Combining two-band $|d_2|^2$ (blue dashed) and three-band $|d_3|^2$ terms (red solid) calculated from perturbation theory into $|d_2+d_3|^2$ (green solid) correctly matches the result from finite displacements (filled light green). Only the three-band plot is to be compared with experiments: Raman intensity is suppressed at the A/B excitons and amplified at the C exciton. The lower panel shows A/B exciton eigenvalues far below all others and eigenvalues near the C exciton bunched together.
}
\label{comp}
\end{figure}

We first explain the detailed theoretical and computational basis of the calculations; readers interested primarily in the results and physical interpretation of excitonic effects in resonant Raman spectra may advance to Fig.~\ref{comp} and the associated discussion.
First-principles Raman spectra calculations are most straightforward for Raman shifts, routinely achieving excellent agreement with experiments \cite{Rice2013, Liang2014, Molina-Sanchez2011}. Raman \textit{intensities} are  usually computed within the non-resonant Placzek approximation: since the scattered light intensity is proportional to the electronic susceptibility $\chi(\omega)$ periodically modified by atomic vibrations ($\omega$ is the incident light frequency), a product-to-sum identity converts the scattered $\cos(\omega_\text{phonon}t) \cos(\omega t)$ wave into Stokes and anti-Stokes components. The scattering amplitude depends on how strongly $\chi$ is changed by vibrations $\xi$, i.e. $|d\chi/d\xi |^2 \propto |d\epsilon/d\xi |^2 $, where $\epsilon(\omega)$ is the dielectric function $\epsilon(\omega) = 1+\sum_S \mesq{0}{\textbf{r}}{S} /(\omega_S-\omega-i\gamma)$ and $S$ runs over all excitations (the ``negative frequency'' contribution is suppressed for clarity but is included in all calculations). This derivative has been calculated using the second derivative of the electronic density matrix \cite{Lazzeri2003, Wirtz2005},  the ``2n+1'' theorem \cite{Veithen2005}, or finite differences of the static dielectric tensor \cite{Umari2001, Liang2014} (calculated from density functional perturbation theory \cite{Baroni2001}). The derivative can also be expanded using perturbation theory, i.e. by treating $\omega_S$ and the matrix elements separately,
\begin{equation}
\label{eq:d2d3}
\begin{split}
\!\!\!\frac{d\epsilon(\omega)}{d\xi} \!&= \sum_S\!
\left(
\frac{\partial\epsilon}{\partial\omega_S }
\frac{d\omega_S}{d\xi} +
\frac{1}{\omega_S-\omega-i\gamma} \frac{d \mesq{0}{\textbf{r}}{S} }{d\xi}
\right) \\
&\equiv d_2 + d_3.
\end{split}
\end{equation}
The former group of ``two-band terms"
\begin{equation}
\label{eq:d2general}
d_2=-\sum_S \frac{\me{0}{\textbf{r}}{S} \me{S}{\partial H}{S} \me{S}{\textbf{r}}{0}}{(\omega_S-\omega-i\gamma)^2} 
\end{equation}
involves only transitions between pairs of bands. The latter group of ``three-band terms'' (see Supplemental Materials)
\begin{equation} \label{eq:d3general}
\begin{split}
\!\!d_3=&\!\sum_{S'\neq S} \frac{-1}{\omega_S\!-\!\omega\!-\!i\gamma} 
\!\left( 
\frac{\me{0}{\textbf{r}}{S'}\me{S'}{\!\partial H}{S} \me{S}{\textbf{r}}{0}}{\omega_{S'}-\omega_S}
\right)\\
+&\!\sum_{S\neq S'} \!\frac{1}{\omega_{S'}\!-\!\omega\!-\!i\gamma} 
\!\left(
\frac{\me{0}{\textbf{r}}{S'}\me{S'}{\!\partial H}{S} \me{S}{\textbf{r}}{0}}{\omega_{S'}-\omega_S}
\right)\\
=&\sum_{S\neq S'} -\frac{\me{0}{\textbf{r}}{S'}\me{S'}{\!\partial H}{S} \me{S}{\textbf{r}}{0}}
{(\omega_{S'}-\omega-i\gamma)(\omega_S-\omega-i\gamma) }\\
\end{split}
\end{equation}
contains transitions between three states. So far the expressions are general: if all quantities are calculated at the DFT level, the Hamiltonian $H=H^\text{DFT}$ and $\kt{S}$ are free electron-hole transitions separated by $\omega_S$; if calculated at the BSE level, $H=H^\text{BSE}$ and $\kt{S}$ are excitonic wavefunctions with eigenvalues $\omega_S$. Physically, two- and three-band terms respectively represent contributions from the oscillating excitation \textit{eigenvalues} (Kohn-Sham eigenvalues or BSE eigenvalues) and the oscillatory \textit{rehybridization} of wavefunctions (Kohn-Sham orbitals or BSE eigenvectors) \cite{Compaan1984}. By combining the $d_2$ and $d_3$ terms we recover the usual perturbation expression for Raman susceptibility $\alpha$, ignoring small phonon energies,
\begin{equation}\label{eq:pert}
\alpha_\text{perturb.} \propto \sum_{S,S'} \frac{\me{0}{\textbf{r}}{S'}\me{S'}{\partial H}{S} \me{S}{\textbf{r}}{0}}
{(\omega_{S'}-\omega-i\gamma)(\omega_S-\omega-i\gamma) }.
\end{equation}
Three-band terms are often neglected due to the apparent squared denominator of the two-band terms (see Eqns.~\ref{eq:d2d3} and \ref{eq:d2general}) \cite{DelCorro2016}; the final expanded expression shows that three-band terms become important when the intervals between excitation energies are small. 

So long as laser energies $\omega$ are away from excitation levels so that $\omega_{\text{phonon}} \ll |\omega-\omega_S+i \gamma|$, the Placzek approximation holds \cite{Profeta2001} and finite-displacement calculations using static dielectric tensors \cite{Liang2014, Puretzky2015} agree qualitatively with Raman intensities measured at finite (but sub-bandgap) $\omega$, 
due to the near-constant dielectric function in this regime. The use of Placzek approximation in the resonant regime \cite {Ambrosch-Draxl2002a, Gillet2013} was argued to be problematic in Ref.~\cite{Profeta2001}, where a more rigorous expression is derived that is  equivalent to keeping only the three-band terms $d_3$. These $d_3$ terms correspond to the so-called ``Albrecht B/C terms'' (or Herzberg-Teller terms) in the vibronic theory for resonant Raman intensities in molecules accounting for nuclear wavefunctions \cite{Albrecht1961, Bozio1989, Walter2018}. The seemingly missing  ``Albrecht A terms'' (or Condon terms) \cite{Walter2018} only arise for excitations with finite Frank-Condon shifts and is negligible for delocalized vibrations in solids \cite{CardonaBook2, Kurti1991} (and even for localized vibrations near certain common defects in MoS$_2$ \cite{Gupta2018}). 

Since $d_3$ readily separates from $d_2$ in the perturbation approach, we derive the single-particle expansion for both at the BSE level and numerically verify that their sum matches the spectra obtained from finite displacements within the Placzek approximation and that, for $\omega \rightarrow 0$, $d_3$ (general) and $d_2+d_3$ (Placzek) converge to the same value, i.e. $d_2$ goes to zero. With the optical matrix elements in Eqn.~\ref{eq:pert} readily available in existing GW-BSE codes, we focus on evaluating the exciton-phonon coupling matrix elements. For the derivative of the exciton Hamiltonian $\partial H^\text{BSE}$ within the Tamm-Dancoff approximation, we neglect the contribution from the derivative of the BSE kernel $\partial K$ \cite{strubbe2012}, neglect the derivative of the quasiparticle correction by using $\partial H^\text{QP} \approx \partial H^\text{DFT}$ (as validated in Refs.~\cite{Ismail-Beigi2003, strubbe2012}) so that
\begin{equation} \label{eq:d2}
\begin{split}
d_2 =&  \sum\limits_{S,vck} \frac{  |\!\me{0}{\textbf{r}}{S}\!|^2 \;\; |\!\brkt{S}{vck}\!|^2}{(\omega_S-\omega-i\gamma)^2} \\
&\times\bigg[\me{ck}{\!\partial H^\text{DFT\!}}{ck} \!-\! \me{vk}{\!\partial H^\text{DFT\!}}{vk} \bigg] 
\end{split}
\end{equation}
Here we neglect $c\neq c'$ and $v\neq v'$ terms in Ref.~\cite{strubbe2012} (DFT-level ``three-band'' terms) since they only contribute significantly when the energy separation between bands is similar to phonon energies; for the low-energy electronic structure of MoS$_2$, most band-pairs of small separation are up-down spin copies forbidding interband scattering, with the exception of the valence band top being split by spin-orbit interaction. Although in general bands split by spin-orbit coupling allow interband scattering (yielding significant DFT-level ``three-band'' terms \cite{Compaan1984, Renucci1975}), the spin-orbit Hamiltonian near the valleys in MoS$_2$ only involves $\sigma_z$ so that spins components are decoupled \cite{Xiao2012, Qiu2013}. 
This approximation is numerically justified later. The $d_3$ terms involve
\begin{equation}
\begin{split}
&\me{S'}{\partial H^\text{BSE}}{S} \approx \sum_{vv'cc'k} \brkt{S'}{vck} \brkt{v'c'k}{S} \\
\times&\bigg[ \me{ck}{\partial H^\text{DFT}}{c'k}\delta_{vv'} - \me{v'k}{\partial H^\text{DFT}}{vk}\delta_{cc'}\bigg].
\end{split}
\end{equation}
Again neglecting $c\neq c'$ and $v\neq v'$ terms and substituting into Eqn.~\ref{eq:d3general} gives
\begin{equation}
\begin{split}
\label{eq:d3}
d_3=& \sum_{S\neq S',vck} \frac{
 \me{0}{\textbf{r}}{S'}\me{S}{\textbf{r}}{0}  S_{vck} S'^*_{vck} }{(\omega_S-\omega-i\gamma)(\omega_{S'}-\omega-i\gamma)}\\
&\times\bigg[ \me{ck}{\partial H^\text{DFT}}{ck}- \me{vk}{\partial H^\text{DFT}}{vk}\bigg] .\\
\end{split}
\end{equation}
All calculations will follow Eqns.~\ref{eq:d2} and \ref{eq:d3}.

All GW-BSE calculations are performed using the BerkeleyGW package \cite{Deslippe2012, Rohlfing2000} based on Kohn-Sham eigenvalues and orbitals obtained within the local density approximation, using Quantum ESPRESSO \cite{Giannozzi2009}.  An energy cutoff of 24~Ry, 500 empty bands, and a $12\times12\times1$ k-point grid was used for the dielectric matrix and quasiparticle self-energy, where the Coulomb interaction is truncated in the out-of-plane direction \cite{Ismail-Beigi2006}. The static remainder technique \cite{Deslippe2013} accelerates convergence of the quasiparticle gap. BSE matrix elements are assembled using 3 valence bands and 4 conduction bands on the same grid and interpolated onto a $40\times40\times1$ grid for diagonalization (Haydock iteration is not used because BSE eigenvectors are needed). The Supplemental Material contains details on convergence tests for the above parameters and all calculations involving phonons. Finally, summation over $S\neq S'$ terms are limited to eigenvalue pairs no further apart than 0.3~eV; exciton pairs separated further contribute negligibly due to large denominators in Eqn.~\ref{eq:d3} and their constituent single-particle transitions being from different bands. Increasing this convergence parameter to 0.4 eV changes Raman intensities by at most 2\% (for any laser frequency). We include 800 excitonic states to converge Raman intensities within the 0\textendash3.5~eV spectral range.

\begin{figure}[h]
\centering
\includegraphics[width=0.5\textwidth]{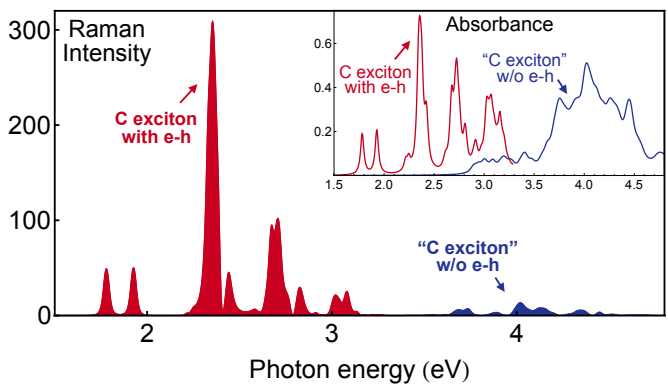}
\caption{Raman intensities with (red) and without (blue) excitonic effects, showing the amplified Raman response at the C exciton compared with the Raman intensities calculated without excitonic effects. The inset compares the absorbance spectra with (red, truncated within its range of convergence) and without (blue) electron-hole interaction, where excitonic effects \textit{redistribute} spectral weights without enhancement. Both Raman intensities shown are from finite displacements; the visible A/B resonances here should be suppressed in the more rigorous three-band spectra ($|d_3|^2$ in Fig.~\ref{comp}).}
\label{regulate}
\end{figure}

The calculated Raman intensities $|\alpha(\omega)|^2$ for the out-of-plane $A'_1$ mode in Fig.~\ref{comp} shows that combining $|d_2|^2$ (blue dashed) and $|d_3|^2$ terms (red solid) from the perturbation approach into $|d_2+d_3|^2$ (green solid) yields good agreement with the finite displacement spectrum (filled green) from pre-resonance ($<$1.5~eV) well into the resonant regime, and that two-band terms correctly converge to zero for vanishing laser energies. These agreements are absolute, i.e. with no adjustable rescaling parameter. While the exclusion of  $c\neq c'$ and $v\neq v'$ terms (DFT-level three-band terms) proved valid, $S\neq S'$ terms (BSE level three-band terms) contribute significantly near the C exciton energy $\sim$2.4~eV. Optical transitions within the near-parallel valence and conduction bands along $\Gamma\!-\!K$ (see band structure in the Supplemental Material) yield a peak in the joint density of states and hence also in the absorbance spectra, ignoring excitonic effects, near 4~eV (blue hollow in inset of Fig.~\ref{regulate}). Including excitonic effects, these transitions are constituents of the C excitons with BSE eigenvalues bunched near 2.4~eV \cite{Qiu2013} (red hollow in Fig.~\ref{regulate}, truncated within its range of convergence). This bunching does \textit{not} cause an order-of-magnitude change in the absorbance spectral features (whose integral is constrained by the f-sum rule \cite{Hybertsen1986}), apart from an overall redshift due to the exciton binding energy and a redistribution of spectral weight rendering exciton resonances sharper than single-particle features. 
However, as in standard perturbation theory where smaller eigenvalue intervals lead to wavefunctions being more strongly perturbed, bunched BSE eigenvalues cause \textit{strong rehybridization of excitonic states during atomic vibration} (i.e. decreased denominators $\omega_{S'}-\omega_S$ in the first line of Eqn.~\ref{eq:d3general}) and regroups what used to be independent transitions at different k-points (which cannot scatter into each other by a $\Gamma$ phonon) into excitonic states all with zero momenta (which allows inter-scattering i.e. increased numerator in Eqn.~\ref{eq:d3}). Therefore, three-band terms contribute an order-of-magnitude amplification in Raman intensities around the C exciton resonance. This can be seen even in the results from finite displacements in Fig.~\ref{regulate}, where Raman intensities without electron-hole interaction near 4~eV (blue filled) are amplified to form the highest Raman peak with electron-hole interaction near 2.4~eV (red filled); comparing the more rigorous three-band spectra would yield the same conclusion.
In stark contrast, the A and B excitons -- each doubly degenerate (two valleys) -- are well separated from other excitations, so they only contribute to two-band terms (dashed blue in Fig.~\ref{comp}). Since only three-band terms are valid for on-resonance frequencies, the orphaned A and B states should not appear in an experimental measurement. Thus the final frequency-dependent Raman intensity $|d_3|^2$ (red in Fig.~\ref{comp}) is suppressed at the A/B excitons and amplified at the C exciton. In this way, our perturbation method reveals how spectral features in resonant Raman characterize not only the exciton spectrum and wavefunction character, but also how exciton-phonon coupling enables inter-state scattering.

\begin{figure}[h]
\centering
\includegraphics[width=0.5\textwidth]{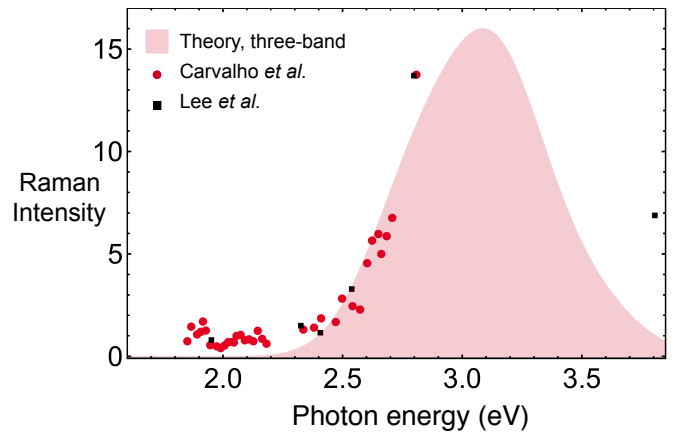}
\caption{Experimental Raman excitation profile for the out-of-plane $A'_1$ mode from Ref.~\cite{Carvalho2016} (red) and Ref.~\cite{Lee2015} (black), compared with the calculated three-band terms in Fig.~\ref{comp} with broadening increased to 0.2~eV to reflect more realistic exciton lifetimes as estimated from free electron lifetimes.}
\label{exp}
\end{figure}

We now compare with experiments in Fig.~\ref{exp} and demonstrate that agreement is only achieved for the beyond-Placzek treatment of Raman intensity including excitonic effects. Two sets of experimental data on the frequency-dependent $A'_1$ mode intensity from Refs.~\cite{Carvalho2016, Lee2015} are aligned at the 2.8~eV data point and normalized in intensity by the Raman peak of silicon at 520~cm$^{-1}$ (which has its own known frequency dependence) to yield the modulus-squared of the Raman susceptibility $|\alpha(\omega)|^2$ (to be distinguished from Raman cross-section, which has an additional $\omega^4$ frequency dependence \cite{Renucci1975, Compaan1984}), which can be directly compared with the calculated results. The calculated three-band intensity from Fig.~\ref{comp} is broadened by 0.2~eV to reflect more realistic C exciton lifetimes estimated from those of free carriers in MoS$_2$ \cite{Li2013}. Good agreement is achieved for the Raman intensity suppression around the A and B excitons, as clearly resolved by the red points (not missing potential resonances) and for the Raman intensity amplification near the C exciton. The two very small resonances measured at the A/B exciton energies and a scissors shift applied are discussed in the Supplemental Materials. In all prior comparisons between finite displacement BSE calculations (Placzek) and experiments known to us, satisfactory agreements were achieved for few laser frequencies \cite{Miranda2017} or for limited spectral region (e.g. the lowest excitonic peak in \cite{Gillet2013}, WS$_2$ A/B excitons in \cite{DelCorro2016}, and the WSe$_2$ C exciton in \cite{DelCorro2016}). Going beyond Placzek allows us to achieve agreement over the energy range of all three excitons.

This analysis has broader implications. For the band structure of a generic solid, every exciton bound state from the solution of the BSE consists of electron-hole pairs with matching group velocities, either at band extrema (zero velocity, spanning a direct gap) or along parallel bands (finite velocity, more common in indirect band gap materials). We expect band-extrema excitons in general to suppress Raman response: by construction these excitons have energies well below parallel-band excitons, giving large denominators in Eqn.~\ref{eq:d3}. Even when there are multiple degenerate valleys as in the case of MoS$_2$, the localized (in k-space) nature of band-extrema excitons allows us to approximate the electron-phonon coupling matrix elements to be constants in Eqn.~\ref{eq:d3}, so (focusing on one $vc$ pair) the sum $\sum_{k} \brkt{S'}{k} \brkt{k}{S}$ can be contracted to zero due to the orthogonality of $S$ and $S'$, giving a vanishing numerator. By contrast, we expect parallel-band excitons in general to amplify Raman response: 
by construction, parallel pairs of conduction and valence bands span larger Brillouin zone areas (often emanating from high symmetry points, which gives them a further multiplicative degeneracy factor) and therefore allow abundant ways of assembling into excitons with similar energies bunching in a narrow energy window (as many as there are sampled k-points in the parallel-band areas). The resonant Raman intensity of silicon amplified by excitonic effects (compared with the independent quasiparticle case) in Ref.~\cite{Gillet2013} is presumably attributed to this mechanism, given the abundance of parallel bands in silicon \cite{Chelikowsky1974}. As a consequence of the general validity of the three-band dominance demonstrated here, resonant Raman measurements can directly probe how excitons undergo \textit{inter}-state scattering by phonons, which affects exciton population dynamics and lifetimes \cite{Antonius2017}. 
 
The perturbation framework developed here not only allows us to go beyond the classical Placzek approximation and include excitonic effects, but also to achieve better scaling behavior: the GW-BSE routine is only performed once statically (at the slight expense of calculating electron-phonon coupling matrix elements for all Raman active modes), compared with finite differences methods where at least two GW-BSE runs are needed for each Raman active mode. This advantage can be exploited to accelerate Raman intensity calculations for low-symmetry materials such as ReS$_2$ \cite{McCreary2017}, with 18 Raman modes.
For second-order Raman intensities, the computational demand for finite differences is even higher, requiring evaluating the BSE dielectric function $N^2$ times, $N$ being the number of Raman modes. In addition, finite-momentum phonon displacements need to be performed on supercells compatible with phonon wavevectors. Despite the computational challenge, second-order Raman intensities were successfully calculated from first-principles recently \cite{Gillet2017}. Our perturbation treatment can be naturally extended to calculate second-order Raman, where the electron-phonon coupling matrix elements would also be calculated for finite-momenta phonons, but without employing supercells thanks to density functional perturbation theory. The key challenge would be in efficiently calculating finite momentum excitons \cite{Cudazzo2016, Qiu2015} (exciton dispersions), which may be overcome using accurate tight-binding based models (fitted to GW band structures) \cite{Wu2015a}. 

This work is supported by computational time on the LSU-superMIC through the XSEDE allocation TG-DMR170050 and by the National Science Foundation Materials Innovation Platform under DMR-1539916. B.R.C. acknowledges the financial support from the Brazilian agencies CNPq and CAPES.

\bibliography{RR}

\end{document}